\documentstyle[aps,prb,preprint]{revtex}

\def\be{\begin{equation}}
\def\ee{\end{equation}}
\def\ie{{\it i.e.}}
\def\eg{{\it e.g.}}
\catcode `\@=11
\@addtoreset{equation}{section}

\def\m@th{\mathsurround=\z@}
\def\ialign{\everycr{}\tabskip\z@skip\halign} 
\def\eqalign#1{\null\,\vcenter{\openup\jot\m@th
  \ialign{\strut\hfil$\displaystyle{##}$&$\displaystyle{{}##}$\hfil
      \crcr#1\crcr}}\,}
\def\matrix#1{\null\,\vcenter{\normalbaselines\m@th
    \ialign{\hfil$##$\hfil&&\quad\hfil$##$\hfil\crcr
      \mathstrut\crcr\noalign{\kern-\baselineskip}
      #1\crcr\mathstrut\crcr\noalign{\kern-\baselineskip}}}\,}
\catcode`@=12

\def\ld{\mathop{\cdots}\limits}
\def\pois#1#2{\{#1,#2\}}         
\def\dd#1{{\partial \over \partial#1}}
\def\g{{\cal G}}

\begin{document}
\preprint{
\vbox{
\hbox{DAMTP/96--99}
\vglue -0.3cm
\hbox{FTUV/96--75 \quad IFIC/96--84}
\vglue -0.3cm
\hbox{hep-th/9612186}
\vglue -0.3cm
\hbox{(to appear in J. Math. Phys.)}
}}

\title
        {The $Z_2$-graded Schouten-Nijenhuis bracket and generalized
	super-Poisson structures
        \footnote{Running Title:
                                Super-Schouten bracket and super-Poisson 
                                structures.}}

\author{
        J.~A.~de~Azc\'{a}rraga $^1$ 
                \footnote
                        {St. John's College Overseas Visiting Scholar.}
                \footnote
                        {On sabbatical (J.A.) leave and on leave of absence 
                        (J.C.P.B.) from 2 above.}
        J.~M.~Izquierdo $^2$, A.~M.~Perelomov $^2$ 
                \footnote
                        {On leave of absence from Institute for Theoretical 
                        and Experimental Physics, 117259 Moscow, Russia. 
                        Present address: Facultad de C. 
                        F\'{\i}sicas, Departamento de F\'{\i}sica Te\'orica, 
                        Univ. Zaragoza, 50009-Zaragoza, Spain.} 
        and J.~C.~P\'{e}rez~Bueno $^1$ 
                \footnotemark[2]
                        \\[0.5cm]}
\address{
        $^1$ Department of Applied Mathematics \& Theoretical Physics \\
        University of Cambridge, Cambridge CB3 9EW, UK \\[0.4cm]
        $^2$ Departamento de F\'{\i}sica Te\'orica and IFIC, \\
        Centro Mixto Universidad de Valencia-CSIC \\
        E-46100-Burjassot (Valencia) Spain.}

\maketitle

\begin{abstract}
The super or $Z_2$-graded Schouten-Nijenhuis bracket is introduced. 
Using it, new generalized super-Poisson structures are found which are given 
in terms of certain graded-skew-symmetric contravariant tensors $\Lambda$ 
of even order. 
The corresponding super `Jacobi identities' are expressed by stating that 
these tensors have zero super Schouten-Nijenhuis bracket with themselves 
$[\Lambda,\Lambda]=0$.
As a particular case, we provide the linear generalized super-Poisson 
structures which can be constructed on the dual spaces of simple superalgebras
with a non-degenerate Killing metric. The $su(3,1)$ superalgebra is given as 
a representative example.
\end{abstract}

\pacs{}

\section{Introduction}

We devote this paper to the introduction of the $Z_2$-graded (or `super') 
Schouten-Nijenhuis bracket and to its application in the definition of
super-Poisson brackets and structures, old and new,
extending the approach of \cite{APPa} to the $Z_2$-graded case. 
The generalization of the standard Poisson Brackets (PB) and Poisson
structures (PS) proposed in \cite{APPa} is 
different from that originally given by Nambu \cite{Na} some
twenty years ago, later also considered in \cite{BF,MS} and
further extended by Takhtajan \cite{Ta} (see \cite{NOS} for a
comparison). It is based on 
the consideration of the Schouten-Nijenhuis bracket (SNB) 
\cite{Sc,Ni} which for the standard PS expresses \cite{Lich,Tu}
the Jacobi condition by requiring zero SNB, $[\Lambda,\Lambda]=0$, for the
(skew-symmetric) bivector field $\Lambda$ defining the PS on the manifold $M$.
The generalized Poisson structures (GPS) in \cite{APPa} are then 
defined by skew-symmetric contravariant tensors of {\it even} order 
(even multivectors) $\Lambda^{(2p)}$. In this way, the skew-symmetry of the 
generalized Poisson bracket (which involves $2p$ functions) and the Leibniz 
rule are automatically incorporated by the properties of 
$\Lambda^{(2p)}\in\wedge^{2p}(M)$ ({\it odd} GPS have been introduced
in \cite{NOS}). The generalized Jacobi identity (GJI) is 
now geometrically expressed as 
$[\Lambda^{(2p)},\Lambda^{(2p)}]=0$, and is different from Takhtajan's 
`fundamental identity' which expresses the fact that the time derivative (the 
`adjoint' map) is a derivation of the $n$-bracket \cite{Ta}.

Graded Poisson structures have been considered before \cite{Leites,Kuper} 
(see also \cite{JM,JG} and references therein).
In \cite{Kras} they were called supercanonical structures $S$, where the 
two-vector $S$ was defined as having a vanishing graded Schouten bracket with 
itself.
Poisson supermanifolds (see, \eg\ \cite{CI} and references therein)
require the replacement of the differentiable
manifold $M$ by a supermanifold $\Sigma$.
By `supermanifold' we understand here a finite dimensional topological
space which has locally the structure of a {\it superspace} \ie, a
space the coordinates of which are given by the even and odd elements of a
Grassmann algebra. We thus follow the `geometric' (see \cite{Rogers,dW}) 
rather that the `algebraic' (see \cite{Berezin,Kostant}) approach
(see \cite{Batchelor} for a comparison).
The algebra of functions ${\cal F}(\Sigma)$ on $\Sigma$, endowed with a 
suitable Poisson bracket, becomes a Poisson superalgebra (see, \eg, \cite{KS}). 
In order to generalize the ($Z_2$-graded or) super-Poisson structures (SPS), 
it is convenient to introduce them through a 
$Z_2$-graded skew-symmetric contravariant tensor field of order two, 
or {\it superbivector},
and to express the super-Jacobi identity as the vanishing of a previously
defined $Z_2$-graded SNB for $Z_2$-graded multivectors or {\it
supermultivectors}. 
There are various types of algebras and brackets related to the original 
Schouten \cite{Sc} and Nijenhuis constructions for multivector fields and 
differential forms \cite{Ni,FN} (see \cite{YK} and references therein for a 
discussion of various algebras). Here we shall consider only the mentioned 
case of the SNB for supermultivectors, and will not discuss other graded
constructions as \eg, those using vector-valued forms \cite{LMS}. 
Thus, we shall start by introducing in Sec. II
the super SNB for supermultivector fields. 

Using the result of Sec. II, an outline of super-Poisson 
structures is presented in Sec. III from the $Z_2$-graded SNB point of view.  
Linear super-Poisson structures are introduced at the end of Sec. III; 
they are defined, as their standard (bosonic) counterparts, by the structure 
constants defining the corresponding Lie superalgebra. 
Generalized super-Poisson structures (GSPS) are defined in Sec. IV, and then 
the linear case is considered in Sec. V. In the standard bosonic case it is not
difficult \cite{APPa} to provide (an infinite number of) examples of 
(linear) GPS using the cohomology properties \cite{CE} (see also \eg, \cite{AI}) 
of simple Lie algebras, and in fact these properties may be used to classify 
the linear GPS which may be constructed on them. Simple superalgebras,
however, may have \cite{Kac} a vanishing Killing form. 
In the nondegenerate case, nevertheless, the arguments are similar to those of 
the standard case, and turn out to be related to cohomology of Lie 
superalgebras. 
This is discussed in Sec. V; an example, that of $su(3,1)$, is given 
in Sec. VI, although its presentation makes it directly applicable to other 
superalgebras.

\section{The $Z_2$-graded Schouten-Nijenhuis bracket}

Before giving the local expression for the $Z_2$-graded SNB, it is 
convenient to establish the conventions. Let $\{ x^i\}$ be local 
coordinates on a supermanifold $\Sigma$ and let $\alpha(i)\equiv\alpha(x^i)$
be the $Z_2$-grade or Grassmann {\it parity} [$0$ (even or `Bose') or $1$ 
(odd or `Fermi')] of $x^i$.
Let $\partial_i\equiv\partial/\partial x^i$ and $dx^i$ be derivatives 
and one-forms. Then, 
\be
      \partial_i dx^j=\delta^j_i\ , \ df=dx^i\partial_i f \ ,\
  (X_1\otimes...\otimes X_p)(\omega^1,...,\omega^p)=(-1)^{\Delta_p(\omega,X)}
      X_1(\omega^1)...X_p(\omega^p)\ ,                     \label{octa}
\ee
where $\delta^j_i$ is the usual Kronecker symbol, $X_i$ and $\omega^j$ are 
one-vectors and one-forms respectively, and \cite{dW}
\be
      \Delta_p(\omega,X)=\sum^p_{\scriptstyle r,s=1 \atop \scriptstyle r<s}
        \alpha(\omega^r)\alpha(X_s)\quad .
                                                            \label{octb}
\ee
where $\alpha(\omega),\alpha(X)$ are the grades of $\omega$ and $X$ 
respectively.
For the wedge product we shall use 
\be
  \partial_i\wedge\partial_j\equiv \partial_i\otimes\partial_j-
(-1)^{\alpha(i)\alpha(j)}\partial_j\otimes\partial_i         \label{ext}
\ee
(similarly for $\omega^i$). 
Eq. (\ref{ext}) may be expressed as $\partial_i\wedge\partial_j=
{\tilde \epsilon}_{ij}^{kl}\partial_k\otimes\partial_l$, where
${\tilde \epsilon}_{ij}^{kl}=- (-1)^{\alpha(k)\alpha(l)}
{\tilde \epsilon}_{ij}^{lk}$.

The general `exterior' product may be defined if we now introduce 
the tilded ($Z_2$-graded) Levi-Civita symbol
${\tilde \epsilon}^{i_1...i_{p}}_{j_1...j_p}$.
We define ${\tilde \epsilon}$ by
\be
     {\tilde \epsilon}^{i_1...i_p}_{j_1...j_p}=\sum^{p}_{s=1}(-1)^{s-1}
     (-1)^{\alpha(i_s)[\alpha(i_1)+...+\alpha(i_{s-1})]}\delta^{i_s}_{j_1} 
     {\tilde\epsilon}^{i_1...{\widehat i_s}...i_{p}}_{j_2\;...\;j_p}
     \quad ,\quad 
     {\tilde\epsilon}^i_j\equiv\delta^i_j\quad \label{IVb}
\ee
or, equivalently, by
\be
{\tilde \epsilon}^{i_1...i_p}_{j_1...j_p}=\sum^{p}_{s=1}(-1)^{s-1}
     (-1)^{\alpha(j_s)[\alpha(j_1)+...+\alpha(j_{s-1})]}\delta^{i_1}_{j_s}
     {\tilde\epsilon}^{i_2\;...\;i_{p}}_{j_1...\widehat{j_s}...j_p}\quad .
     \label{IVbb}
\ee
Clearly, $\tilde\epsilon$ reduces to the standard $\epsilon$ in the Bose
case ($\alpha(i)=0$) for which
\be
\epsilon^{i_1\ldots i_{p}}_{j_1\ldots j_{p}}=
\left|\matrix{\delta^{i_1}_{j_1} & \cdots & \delta^{i_1}_{j_p} \cr
\vdots & & \vdots \cr
\delta^{i_p}_{j_1} & \cdots & \delta^{i_p}_{j_p} \cr}\right|
\quad.
\label{ebose}
\ee
Expressions (\ref{IVb}) and (\ref{IVbb}) correspond, respectively, to expanding this 
determinant by columns/rows in such a way that the column $(j)$/row $(i)$ 
indices are written in natural order, and then adding the sign factors 
which would be needed to bring the row $(i)$/ column $(j)$ indices to the 
ordering in which they actually appear.
The $Z_2$-graded $\tilde\epsilon$ has the graded antisymmetry property
\be
   {\tilde \epsilon}^{i_1\;...\;i_p}_{j_1...j_kj_{k+1}...j_p}=
   -(-1)^{\alpha(j_k)\alpha(j_{k+1})}
   {\tilde \epsilon}^{i_1\;...\;i_p}_{j_1...j_{k+1}j_k...j_p}    
   \label{IVc}
\ee
and similarly for superscripts. 
As in the even case, we have
\be
\tilde\epsilon^{i_1\ldots i_{p-1} l_p \ldots l_n}_{j_1\quad \ldots\quad j_n}
\tilde\epsilon^{i_p \ldots i_n}_{l_p\ldots l_n}=
(n-p+1)!\;\tilde\epsilon^{i_1\ldots i_n}_{j_1\ldots j_n}
\quad.
\label{econtr}
\ee

Using (\ref{IVc}) we may now write
\be
    \partial_{i_1}\wedge...\wedge\partial_{i_p}:=
    {\tilde \epsilon}_{i_1...i_p}^{j_1...j_p}\partial_{j_1}\otimes...
  \otimes\partial_{j_p} \quad .                                
   \label{extproddef}
\ee
Then, the exterior product of two general
supermultivectors $A$ and $B$ of parities $\alpha(A)$, $\alpha(B)$ and of order 
$p$ and $q$, respectively, locally expressed by
\be
      A={1\over p!}A^{i_1...i_p}\partial_{i_1}\wedge...\wedge\partial_{i_p}
      \quad,\quad
      B={1\over q!}B^{j_1...j_q}\partial_{j_1}\wedge...\wedge\partial_{j_q}
     \quad ,                                             \label{appone}
\ee
is given by
\be
\eqalign{
 A\wedge B\equiv 
 &
 {1\over p!q!}(-1)^{[\alpha(i_1)+...+\alpha(i_p)]
 [\alpha(j_1)+...+\alpha(j_q)+\alpha(B)]}\ \cdot
 \cr
 \cdot\, &
 A^{i_1...i_p}B^{j_1...j_q}
 \partial_{i_1}\wedge...\wedge\partial_{i_p}\wedge
 \partial_{j_1}\wedge...\wedge\partial_{j_q}
 \cr
 =&
 {1\over (p+q)!}(A\wedge B)^{i_1\ldots i_p j_1\ldots j_q}
\partial_{i_1}\wedge...\wedge\partial_{i_p}\wedge
\partial_{j_1}\wedge...\wedge\partial_{j_q}\quad,\cr}    
\label{octc}
\ee
where
\be
(A\wedge B)^{i_1\ldots i_p j_1\ldots j_q}=
{1\over p!q!}(-1)^{[\alpha(k_1)+...+\alpha(k_p)]
[\alpha(l_1)+...+\alpha(l_q)+\alpha(B)]}
\tilde\epsilon^{i_1\ldots i_p j_1\ldots j_q}_{k_1\ldots k_p l_1\ldots l_q}
A^{k_1...k_p}B^{l_1...l_q}
\quad.\label{coord}
\ee
The contravariant exterior algebra of supermultivectors on $\Sigma$ will be 
denoted by $\wedge(\Sigma)$; a vector field $X$ belongs to 
$\wedge^1(\Sigma)$ and 
${\cal F}(\Sigma)=\wedge^0(\Sigma)$. 
Due to the $Z_2$-gradation (see (\ref{ext})), there may exist supermultivectors of 
arbitrary order $p$ irrespective of the dimension of $\Sigma$ if $\Sigma$ 
is not an ordinary manifold.

We are now in position to generalize the Schouten-Nijenhuis bracket to the 
supersymmetric case.
The way to proceed is the following.
On an ordinary manifold $M$ it is known that the SNB of 
$X_1\wedge\ldots\wedge X_p$ and $Y_1\wedge\ldots\wedge Y_q$, 
$X_i,Y_j\in\wedge^1(M)$, is the bilinear local operation given by
\be
\eqalign{
[X_1\wedge...\wedge X_{p}\ ,&
\ Y_{1}\wedge \ldots \wedge Y_{q}] =
\cr
= \sum^{p,q}_{s=1,t=1} (-1)^{t+s}X_{1}\wedge...\hat X_{s}...&
\wedge X_{p}
\wedge[X_{s} ,Y_{t}]\wedge Y_{1} \wedge ... \hat Y_{t} ...
\wedge Y_{q}\quad.
\cr}
\label{snbb}
\ee
This formula is the result of extending the Lie derivative of vector fields 
on $M$, $L_X Y=[X,Y]$, in a natural (and unique) way to arbitrary elements of 
the contravariant exterior algebra $\wedge(M)$.
Now, substituting $\Sigma$ for $M$, it is not difficult to see that for 
$X_i,Y_j\in\wedge^1(\Sigma)$ (\ref{snbb}) is replaced by
\be
\eqalign{
&
[X_1\wedge...\wedge X_{p}\ , 
\  Y_{1}\wedge \ldots \wedge Y_{q}] =
\cr
= \sum^{p,q}_{s=1,t=1} & 
(-1)^{t+s+\alpha(Y_t)[\alpha(Y_1)+\ldots+\alpha(Y_{t-1})]
+\alpha(X_s)[\alpha(X_{s+1})+\ldots+\alpha(X_{p})]} 
\cr 
& X_{1}\wedge...\hat X_{s}... 
\wedge X_{p}
\wedge[X_{s} ,Y_{t}]\wedge Y_{1} \wedge ... \hat Y_{t} ...
\wedge Y_{q}
\cr}
\label{ssnba}
\ee
where in the $r.h.s$ $[\ ,\ ]$ denotes the $Z_2$-graded SNB of two vector 
fields (which is an anticommutator if both are odd).
Using (\ref{ssnba}) we may now introduce the SSNB for arbitrary elements of 
$\wedge(\Sigma)$.
This leads us to the following local definition:

\medskip
\noindent {\bf Definition II.1} ({\it Super Schouten-Nijenhuis bracket})

\noindent Let  $A\in \wedge^p(\Sigma)$ and $B\in 
\wedge^q(\Sigma)$ be two supermultivectors on a supermanifold $\Sigma$ given 
locally by (\ref{appone}).
The super Schouten-Nijenhuis 
bracket (SSNB) $[A,B]$ is the (super-)bilinear operation of local type 
$[\ ,\ ]:\wedge^p(\Sigma)\times\wedge^q(\Sigma)\to\wedge^{p+q-1}(\Sigma)$ 
locally defined by
\be
     \eqalign{[A,B]&={1\over (p-1)!q!}(-1)^{[\alpha(i_1)+...+\alpha(i_{p-1})]
     [\alpha(j_1)+...+\alpha(j_q)+\alpha(B)]}A^{\nu i_1...i_{p-1}}\cr
     &\partial_\nu
     B^{j_1...j_q}\partial_{i_1}\wedge...\wedge\partial_{i_{p-1}}\wedge
     \partial_{j_1}\wedge...\wedge\partial_{j_q}\cr
      &+ {(-1)^p\over p!(q-1)!}(-1)^{\alpha(A)[\alpha(j_1)+...+\alpha(j_{q-1})
      +\alpha(B)]}B^{\nu j_1...j_{q-1}}\cr
      &\partial_\nu A^{i_1...i_p}\partial_{i_1}
      \wedge...\wedge\partial_{i_p}\wedge\partial_{j_1}\wedge...
      \wedge\partial_{j_{q-1}} \quad ,\cr}                \label{apptwo}
\ee
where $\alpha(i_k)$, $\alpha(j_k)$ ($\alpha (A)$, $\alpha(B)$) are the 
Grassmann parities of the corresponding coordinates (supermultivectors $A$ and 
$B$); $\alpha([A,B])=\alpha(A)+\alpha(B)$.

If we write now 
$[A,B]={1\over (p+q-1)!}[A,B]^{i_1\ldots i_{p+q-1}}
\partial_{i_1}\wedge\ldots\wedge\partial_{i_{p+q-1}}$, the coordinates of the 
SSNB are given by [eq. (\ref{coord})]
\be
   \eqalign{[A,B&]^{k_1...k_{p+q-1}}=
   {1\over (p-1)!q!}(-1)^{[\alpha(i_1)+...+\alpha(i_{p-1})]
     [\alpha(j_1)+...+\alpha(j_q)+\alpha(B)]}\cr
     &{\tilde\epsilon}^{k_1...k_{p+q-1}}_{i_1...i_{p-1}j_1...j_q}
     A^{\nu i_1...i_{p-1}}\partial_\nu B^{j_1...j_q}\cr
     &+{(-1)^p\over p!(q-1)!}(-1)^{\alpha(A)[\alpha(j_1)+...+\alpha(j_{q-1})
      +\alpha(B)]}\cr
      &{\tilde\epsilon}^{k_1...k_{p+q-1}}_{i_1...i_pj_1...j_{q-1}}
      B^{\nu j_1...j_{q-1}}\partial_\nu A^{i_1...i_p}\quad .\cr}
                                                            \label{apptwoa}
\ee

Expression (\ref{apptwo}) reproduces the
definition of the graded commutator when $A$ and $B$ are graded vector fields
on $\Sigma$ ($A,B\in\wedge^1(\Sigma)$). 
It also reduces to the local expression of the usual Schouten-Nijenhuis bracket 
(see, \eg\ \cite{Lich}) for the bosonic case
($\alpha(x^i)=0,\ \alpha(A)=0,\ \alpha(B)=0$) as it should. 
It follows from (\ref{apptwo}) that the SSNB has the following
property ($A\in\wedge^p(\Sigma), B\in\wedge^q(\Sigma)$):
\be
       [A,B]=(-1)^{pq}(-1)^{\alpha(A)\alpha(B)}[B,A] \quad .\label{appthree}
\ee
As a result, $[A,A]$ is identically zero for a Grassmann
even $p$-multivector  if $p$ is {\it odd}, and $[A,A]=0$ 
is a non-trivial equation if $A$ is of zero parity and $p$ is even. 
Also, if $C\in\wedge^r(\Sigma)$
\be
  \eqalign{   (-1)^{\alpha(A)\alpha(C)}&(-1)^{pr}\big[[A,B],C\big] +
  (-1)^{\alpha(B)\alpha(A)}(-1)^{qp}\big[[B,C],A\big]
      \cr
     &+(-1)^{\alpha(C)\alpha(B)}(-1)^{rq}\big[[C,A],B\big]
     =0 \quad ,
     \cr}                                                   \label{appfour}
\ee
\be
     \eqalign{[A,B\wedge C]&=[A,B]\wedge C+ (-1)^{(p-1)q}(-1)^{\alpha(B)
     \alpha(A)}B\wedge [A,C]  \quad ,\cr
     [A\wedge B, C]&=(-1)^pA\wedge [B,C]+(-1)^{rq}(-1)^{\alpha(B)\alpha(C)}
     [A,C]\wedge B \quad .\cr}                                \label{appfive}
\ee
Notice that if $A$ is a vector field, the first in (\ref{appfive}) is just the 
derivation property
\be
     L_A(B\wedge C)=(L_A B)\wedge C + (-1)^{\alpha(A)\alpha(B)}B\wedge(L_A C)
     \quad,
\ee
where $L_A$ is the Lie derivative with respect to $A$.
                                                                 
The dependence on $p,q$ of eqs. (\ref{appthree}) and (\ref{appfour}) indicates 
that the definition of the SSNB in eq. (\ref{apptwo}) does not have the usual 
properties of a superalgebra bracket with respect to the Grassmann parity 
grading. 
This may be achieved if (\ref{apptwo}) is slightly modified and a new grading 
$\pi$ for supermultivectors $A$ is introduced. 
The {\it degree} $\pi(A)\equiv a$ is defined by
\be
    \pi(A):=\alpha(A)+p-1 \quad ,               \label{appfivea}
\ee
where $p$ is the order of the supermultivector $A$. It then follows that
$\pi([A,B]) = \alpha(A)+\alpha(B)+(p+q-1)-1= \pi(A)+\pi(B)\equiv a+b$. 
If we now define a new, primed SSNB by 
\be
    [A,B]{'}=(-1)^{p+1}(-1)^{\alpha(A)(q+1)}[A,B] \quad ,  \label{appsix}
\ee
where $[A,B]$ is the old one given by (\ref{apptwo}), properties (\ref{appthree}) 
and (\ref{appfour}) now adopt the superalgebra form 
(see, {\it e.g.}, \cite{SCHEUNERT,dW})
\be
    [A,B]{'}=-(-1)^{ab}[B,A]{'}\quad ,            \label{appsevena}
\ee
\be
     (-1)^{ac}\big[[A,B]{'},C\big]{'}+(-1)^{cb}\big[[C,A]{'},B\big]{'}
     +(-1)^{ba}\big[[B,C]{'},A\big]{'}=0 \quad ,            \label{appseven}
\ee
where eq. (\ref{appseven}) is the standard super-Jacobi identity; 
$[A,A]$ is trivially zero for $a$ even. 
Thus, the primed SSNB bracket above extends the 
superalgebra of supervector fields $X$ (for which the parity associated 
with $\pi$ is just the
Grassmann one $\pi(X)=\alpha(X)$, eq. (\ref{appfivea})), and makes a superalgebra
of the exterior algebra of supermultivectors endowed with the SSNB.

\section{Super-Poisson structures}

Let $\Sigma=\Sigma_0\oplus \Sigma_1$ be a supermanifold and 
${\cal F}(\Sigma)={\cal F}_0(\Sigma)\oplus{\cal F}_1(\Sigma)$ 
be the algebra of $Z_2$-graded smooth functions on $\Sigma$; 
$f\in{\cal F}_0(\Sigma)$ $[{\cal F}_1(\Sigma)]$ is
said to be homogeneous of even [odd] parity. 

\medskip
\noindent
{\bf Definition III.1} ({\it Super-Poisson bracket})

\noindent A {\it super-Poisson bracket} $\{ \cdot ,\cdot \}$ (SPB) on 
${\cal F}(\Sigma)$ is a
bilinear operation assigning to every pair of functions 
$f, g\in {\cal F}(\Sigma)$ a new function $\{f, g\}\in {\cal F}(\Sigma)$, 
such that for homogeneous functions satisfies the following conditions:

\noindent
a) grade zero super-Poisson bracket
\be
\alpha(\{f,g\})\equiv\alpha(f)+\alpha(g)\ (\hbox{mod}\ 2)\quad ;
\label{suma}
\ee
\noindent
b) super skew-symmetry 
\be
\{f, g\} = - (-1)^{\alpha(f)\alpha(g)}\{g, f\}\; ;\label{pri}
\ee
c) graded Leibniz rule (derivation property)
\be
\{f, gh\} = \{ f, g\} h+ (-1)^{\alpha(f)\alpha(g)}g\{ f,h\} =
\{ f, g\} h+ (-1)^{\alpha(g)\alpha(h)}\{ f,h\}g \quad ,\label{ter}
\ee
d) super-Jacobi identity
\be
\eqalign{
{1\over 2}\hbox{sAlt} \{f_1,\{f_2,f_3\}\}\equiv
&
(-1)^{\alpha(1)\alpha(3)}
\{f_1, \{f_2, f_3\}\} + 
(-1)^{\alpha(2)\alpha(1)}
\{f_2, \{f_3,f_1\}\}\cr
+&
(-1)^{\alpha(3)\alpha(2)}
\{f_3, \{f_1,f_2\}\}= 0\;,\cr}
\label{sec}
\ee
where $\alpha(i)\equiv\alpha(f_i)$, $i=1,2,3$ and sAlt means `super'
or $Z_2$-graded alternation. Since the identities (\ref{pri}), 
(\ref{sec}) are just the axioms of a superalgebra,
the space ${\cal F}(\Sigma)$ endowed with the SPB
$\{ \cdot ,\cdot \}$
becomes an (infinite-dimensional) superalgebra, and $\Sigma$ is a 
{\it super-Poisson space}.
The first of the above conditions means that the SPB operation 
$\{\cdot,\cdot\}$ itself is Grassmann
even. We shall restrict ourselves here to this case although odd PB 
(`antibrackets') for which $\alpha(\{f,g\})=\alpha(f)+\alpha(g)+1$ 
appear in the theory of odd supermechanics \cite{Leites}
(see also \cite{Kuper} and the Remark below).

\medskip
\noindent
{\it Remark.}\quad 
Note the way the grading $[\alpha]$ has been defined. 
Odd structures have also appeared in mathematics in connection with the SNB 
(see \cite{Buttin}) and in physics, as in the Batalin-Vilkovisky formalism 
\cite{BV,Witten,LZ,ZWI,AS}; see also \cite{YK} and references therein and in
\cite{HIGHER}.

Let $x^{j}$ be coordinates on $\Sigma$ and consider SPB
of the form
\be
\{f(x), g(x)\} := (-1)^{\alpha(f)\alpha(k)+\alpha(j)\alpha(k)}
\omega ^{jk}(x)\partial _{j}f\partial _{k}g\quad,\quad 
j,k=1,\ldots, \hbox{dim}\Sigma \quad,
\label{cua}
\ee
where $\alpha(i)$ is as before
and $\partial _{j} = {\partial / {\partial x^{j}}}$ is a left
derivative. Clearly, it satisfies (\ref{pri}). Note that, in particular,
\be
    \{ x^i, x^j \}=\omega^{ij}\quad             \label{cuaa}
\ee
and that we take here 
$\alpha(\omega^{ij})=\alpha(x^i)+\alpha(x^j)$.
Since the graded Leibniz rule is automatically guaranteed by 
(\ref{cua}), $\omega^{ij}(x)$ defines a 
SPB if $\omega ^{ij}(x) = -(-1)^{\alpha(i)\alpha(j)}\omega ^{ji}(x)$
and eq. (\ref{sec}) is satisfied \ie, if
\be
(-1)^{\alpha(j)\alpha(m)}\omega ^{jk}\partial _{k}\omega ^{lm} + 
(-1)^{\alpha(l)\alpha(j)}\omega ^{lk}\partial _{k}\omega^{mj} + 
(-1)^{\alpha(m)\alpha(l)}\omega ^{mk}\partial _{k}\omega ^{jl} = 0\;,
\label{sex}
\ee
which is equivalent (cf. (\ref{sec})) to $s\hbox{Alt}[\omega^{jk}
\partial_k\omega^{lm}]:={1\over 2}\tilde\epsilon^{j\,l\,m}_{i_1 i_2 i_3}
\omega^{i_1 k}\partial_k\omega^{i_2 i_3}=0$.

The requirements (\ref{suma}), (\ref{pri}) and (\ref{ter}) imply that the SPB may be 
given in terms of a $Z_2$-graded bivector field or {\it super-Poisson bivector}
$\Lambda\in\wedge ^2(\Sigma)$ of zero parity. 
Locally,
\be
\Lambda = {1\over 2} (-1)^{\alpha(j)\alpha(k)}
\omega ^{jk}\partial _j\wedge \partial _k= -{1\over 2}\omega^{kj}\partial_j
\wedge\partial_k\quad .\label{nueva}
\ee
Condition (\ref{sex}) may now be expressed in terms of $\Lambda$ and the 
SSNB (eq. (\ref{apptwo})) as $[\Lambda, \Lambda]=0$.
Thus, if $x^j$, $x^k$ are both odd, $\alpha(x^j)=\alpha(x^k)=1$,
$\omega^{jk}$ is symmetric rather than 
antisymmetric.
A super bivector $\Lambda\in\wedge^2(\Sigma)$   
such that $[\Lambda,\Lambda]=0$ defines a {\it super-Poisson structure} on 
$\Sigma$ and $\Sigma$ itself becomes a {\it super-Poisson space}. 
The SPB is then given by 
\be
\{f,g\}=\Lambda(df,dg)\quad,\quad f,g\in{\cal F}(\Sigma)
\quad.\label{oneone}
\ee
Two SPS $\Lambda_1, \Lambda_2$ on $\Sigma$ are  
{\it compatible} if any linear combination of them is again a SPS. 
In terms of the SSNB this means that 
$[\Lambda_1,\Lambda_2]=0.$

Given a bosonic function $H\in {\cal F}_0(\Sigma)$, the supervector field 
$X_H=i_{dH}\Lambda$ (where 
$i_\alpha\Lambda(\beta):=\Lambda(\alpha,\beta)\,,\,\allowbreak\alpha,\beta$ 
one-forms), is called a
{\it super-Hamiltonian vector field} of $H$. From the 
super-Jacobi identity (\ref{sec}) easily follows that 
\be
[X_f,X_H]=X_{\{ f,H\} }\quad.\label{sep}
\ee
Thus, the super Hamiltonian vector fields span a sub-superalgebra of the 
superalgebra 
${\cal X}(\Sigma)$ of all smooth supervector fields on $\Sigma$.
In local coordinates
\be
X_{H} = (-1)^{\alpha(j)\alpha(k)}\omega ^{jk}(x) \partial _{j}H
\partial _k\quad; \quad
X_H.f =\{H,f\}\;.\label{oct}
\ee

A particular case is that of the {\it linear super-Poisson structures}.
A real finite--dimensional superalgebra $\g$ with $Z_2$-graded Lie bracket
$[\, .,.]$ defines in a natural way a SPB
$\{\, .,.\}_\g$
on the dual space $\g^*$ of $\g$.  The natural identification $\g \cong
(\g^*)^*$, allows us to think of $\g$ as a subset of the ring of smooth
functions ${\cal F}(\g^*)$.
Choosing a linear basis
$\{\,e_i\,\}_{i=1}^r$ of $\g$, and identifying its components with linear
coordinate functions $x_i$ on the dual space $\g^*$ by means of $x_i(x) =
\langle x, e_i\rangle$ for all $x\in \g^*$, the
fundamental SPB on $\g^*$ may be defined by 
\be
\pois{x_i}{x_j}_\g = x_k C_{ij}^{k}  \quad,\quad i,j,k=1,\ldots, r=\hbox{dim}
\g\quad,
\label{IVv}
\ee
using that $[e_i, e_j] =  C_{ij}^{k} e_k$, where $C_{ij}^{k}$ are the
structure constants of $\g$. Since these are of even Grassmann parity, 
assumption a) in Def. III.1 tells us that $\alpha(\{ x_i,x_j\})=\alpha(x_k)$
as is indeed the case.
Intrinsically, the SPB $\pois . . _\g$ on 
${\cal F}(\g^*)$ is defined by 
\be
\pois{f}{g}_\g (x) = \langle x,[df(x),dg(x)]\rangle\quad,\quad
f,g\in {\cal F}(\g^*), x\in \g^*\quad;
\label{IVvi}
\ee
locally, 
$[df(x),dg(x)]=(-1)^{\alpha(f)\alpha(j)+\alpha(i)\alpha(j)}
e_kC_{ij}^k {\partial f \over \partial x_i}
{\partial g \over \partial x_j}\ $, 
$\{f,g\}_\g (x)=(-1)^{\alpha(i)\alpha(j)+\alpha(f)\alpha(j)}
\allowbreak
x_kC_{ij}^k {\partial f \over \partial x_i}
{\partial g \over \partial x_j}$.
The above SPB
$\pois . . _\g$ 
will be called a {\it super Lie--Poisson bracket}.
It is associated to a two-supervector field $\Lambda_\g$ on $\g^*$ 
locally defined by 
\be
\Lambda_\g = (-1)^{\alpha(i)\alpha(j)}{1\over 2}x_k C_{ij}^{k}  
  \dd{x_i}\wedge \dd{x_j}
\equiv -{1\over 2}\omega_{ji}\partial^i
  \wedge\partial^j
\label{IVvii}
\ee
(cf. (\ref{nueva})), so that (cf. (\ref{oneone}))
$\Lambda_\g (df, dg ) = \pois{f}{g}_\g$. Since $\alpha(x_i)=\alpha(\partial^i)$,
we see that condition a) in Def III.1 implies that $\alpha(\Lambda_{\cal G})=0$
and that accordingly the degree $\pi(\Lambda_{\cal G})$ of $\Lambda_{\cal G}$
reduces to $(\hbox{order}(\Lambda_{\cal G})-1)=1$.
We conclude by noting that the non-trivial condition $[\Lambda_\g,
\Lambda_\g]=0$ (cf. (\ref{sex})) reproduces the super-Jacobi identity for 
the superalgebra $\g$ written as
\be
{1\over 2}
\hbox{sAlt}(C^\sigma_{i_1\rho}C^\rho_{i_2 i_3}):=
{1\over 2}
{\tilde \epsilon}^{j_1j_2j_3}_{i_1i_2i_3}
C^\sigma_{j_1 \rho}C^\rho_{j_2 j_3}=0\quad.
\label{VIa}
\ee

\section{Generalized super-Poisson structures}
A rather stringent condition needed to define a SPS on a supermanifold is the 
super-Jacobi identity (\ref{sec}); it will be fulfilled if the coordinates of
the super-Poisson bivector
(\ref{nueva}) satisfy (\ref{sex}). This is expressed in a 
geometrical way by the vanishing of the SSNB of $\Lambda\equiv\Lambda^{(2)}$
with itself, 
$[\Lambda^{(2)}, \Lambda^{(2)}]=0$.
So, it seems natural to consider generalizations of the SPS
in terms of $2p$-ary operations determined by Grassmann even
$2p$-supermultivector fields $\Lambda ^{(2p)}$, the case $p=1$ 
being the standard one. Note that, if we relax condition a) in Def. III.1,
we may have odd Poisson brackets and structures, defined in this case by
Grassmann odd $q$-supermultivectors also of odd order (and hence 
also of odd $\pi$-parity,
eq. (\ref{appfivea})) for which $[\Lambda,\Lambda]=0$ (cf. (\ref{appthree})) will be
non-trivial.

Having this in mind, let us introduce first

\medskip
\noindent
{\bf Definition IV.1} ({\it Generalized super-Poisson bracket})

\noindent A generalized super-Poisson bracket (GSPB)
$\{\cdot ,\cdot ,\ldots ,\cdot ,\cdot \}$ 
on a supermanifold $\Sigma$ is a mapping
${\cal F}(\Sigma)\times \ld^{2p}\times{\cal F}(\Sigma)\to {\cal F}(\Sigma)$
assigning a function $\{f_1, f_2,\ldots ,f_{2p}\}$ to every set 
$f_1,\ldots ,f_{2p}\in {\cal F}(\Sigma)$
which is linear in
all arguments and satisfies the following conditions:

\medskip
\noindent 
a) even GSPB
\be
\alpha(\{f_1, f_2,\ldots ,f_{2p}\})\equiv\alpha(f_1)+\ldots+\alpha(f_{2p})\
(\hbox{mod}\ 2)\quad,
\label{fouronea}
\ee

\medskip
\noindent
b) graded skew-symmetry in all $f_j$;

\medskip
\noindent
c) graded Leibniz rule: $\forall f_i,g,h\in {\cal F}(\Sigma),$
\be
\{f_1,\ldots ,f_{2p-1},gh\} = \{f_1,\ldots ,f_{2p-1},g\}h
+(-1)^{\alpha(g)\alpha(h)}\{f_1,\ldots ,f_{2p-1},h\}g\quad;
\label{fourone}
\ee

\medskip
\noindent
d) generalized super-Jacobi identity: $\forall f_i\in {\cal F}(\Sigma),$
\be
\hbox{sAlt}\,\{f_1,f_2,\ldots ,f_{2p-1},\{
f_{2p},\ldots ,f_{4p-1}\}\} = 0 =
\tilde\epsilon^{j_1\ldots j_{4p-1}}_{1\ldots 4p-1}
\{f_{j_1},f_{j_2},\ldots ,f_{j_{2p-1}},\{f_{j_{2p}},\ldots ,f_{j_{4p-1}}\}\}
\,.
\label{fourtwo}
\ee

The property a) indicates that we are again restricting ourselves to a
Grassmann even GSPB. Conditions b) and c) imply that the GSPB is given by a
super skew-symmetric multiderivative, \ie\ by a
Grassmann even $2p$-supermultivector field $\Lambda ^{(2p)}\in 
\wedge^{2p}(\Sigma)$.
Condition (\ref{fourtwo}) will be called the {\it generalized
super-Jacobi identity}; 
for $p=2$ it contains $35$ terms ($C^{2p-1}_{4p-1}$ in the 
general case). It may be 
rewritten as $[\Lambda ^{(2p)}, \Lambda ^{(2p)}]
=0$ where $\Lambda^{(2p)}$ defines a generalized SPS on $\Sigma$. 
In the bosonic case, where $\alpha$ is always zero and sAlt reduces to Alt, 
these 
generalized Poisson structures have been proposed in \cite{APPa}.
Clearly, the above conditions reproduce (\ref{suma})--(\ref{sec}) for
$p=1$. The compatibility condition in Sec. III for $p=1$ may be now extended in
the following sense:
two generalized super-Poisson structures 
$\Lambda ^{(2p)}$ and ${\Lambda}^{(2q)}$ on $\Sigma$
are called {\it compatible} if they `commute' under the SSNB
\ie, $[\Lambda ^{(2p)}, {\Lambda}^{(2q)}]=0$. For the linear case,
our generalized SPS are automatically
obtained from constant supermultivectors of order $2p+1$.

Let $x^j$ be local coordinates on $U\subset \Sigma$. Then the GSPB
has the form
\be
\{f_1(x), f_2(x),\ldots ,f_{2p}(x)\}=(-1)^{\Delta_{2p}(f,j)}
(-1)^{\Delta_{2p}(j,j)}\omega _{j_1j_2\ldots j_{2p}}
\partial^{j_1}f_1\,\partial^{j_2}f_2\,\ldots \,\partial^{j_{2p}}f_{2p}\quad,
\label{fourthree}
\ee
where 
$\Delta_{2p}(f,j)=\sum^{2p}_{r<s}\alpha(f_r)\alpha(j_s)$,
$\Delta_{2p}(j,j)=\sum^{2p}_{r<s}\alpha(j_r)\alpha(j_s)$ (cf. (\ref{octb}))
and $\omega_{j_1j_2\ldots j_{2p}}$ 
are the coordinates of a graded skew-symmetric tensor which satisfies
\be
\hbox{sAlt}\,(\omega _{j_1j_2\ldots j_{2p-1}k}\,\partial ^k\,
\omega_{j_{2p}\ldots j_{4p-1}}) = 0
\label{fourfour}
\ee
as a result of (\ref{fourtwo}) (notice that this would be false for a bracket with 
an odd number of arguments). In terms of an even supermultivectior field of 
order $2p$ the generalized super-Poisson structure is defined by the 
$2p$-vector
\be
\Lambda ^{(2p)} = {1\over {(2p)!}}(-1)^{\Delta_{2p}(j,j)}
\omega _{j_1\ldots j_{2p}}\,\partial^{j_1}
\land \ldots \land \partial ^{j_{2p}}={(-1)^p\over (2p)!}\omega_{j_{2p}...j_1}
\partial^{j_1}\wedge...\wedge\partial^{j_{2p}}
\label{fourfive}
\ee
and, using (\ref{octa})
\be
\Lambda(df_1,\dots,df_{2p})=\{f_1,\dots,f_{2p}\}\quad.
\ee
\medskip
\noindent
{\bf Lemma IV.1}

\noindent The vanishing of the SSNB
$[\Lambda ^{(2p)},\Lambda ^{(2p)}]=0$ reproduces eq. (\ref{fourfour}).

\noindent
{\it Proof}: Let $\Lambda^{(2p)}$ be the $2p$-vector defined in (\ref{fourfive}). 
To show this, it suffices to use (\ref{apptwo}) for the case $A=B=\Lambda^{(2p)}$,
$\alpha(\Lambda^{(2p)})=0$ since we assumed that the Grassmann parity
of the GPB was determined by those of the $2p$ functions involved 
in it only. Then,
\be
     \eqalign {[\Lambda^{(2p)},\Lambda^{(2p)}]&=- {1\over (2p-1)!(2p)!}
     (-1)^{[\alpha(i_1)+...+\alpha(i_{2p-1})][\alpha(j_1)+...+\alpha(j_{2p})]}
     \omega_{ i_1...i_{2p-1}\nu}\cr
     &(-1)^{\Delta_{2p-1}(i,i)+\Delta_{2p}(j,j)}\partial^\nu 
     \omega_{j_1...j_{2p}}\partial^{i_1}\wedge...\wedge
     \partial^{i_{2p-1}}\wedge\partial^{j_1}\wedge...\wedge\partial^{j_{2p}}
      \cr
     &-{1\over (2p-1)!(2p)!}\omega_{j_1...j_{2p-1}\nu}(-1)^{\Delta_{2p-1}
     (j,j)+\Delta_{2p}(i,i)}\cr
     &\partial^\nu
     \omega_{i_1...i_{2p}}\partial^{i_1}\wedge...\wedge\partial^{i_{2p}}
     \wedge\partial^{j_1}\wedge...\wedge\partial^{j_{2p-1}}\cr
     &=-{2\over (2p-1)!(2p)!}(-1)^{\Delta_{4p-1}(i,i)}
     \omega_{i_1...i_{2p-1}\nu}\partial^\nu
     \omega_{i_{2p}...i_{4p-1}}\partial^{i_1}\wedge...
     \wedge\partial^{i_{4p-1}}
     \quad ,\cr}                                            \label{appeight}
\ee
which, since $\Delta_{4p-1}(i,i)$ is invariant under reorderings of the
indices $i_1,...,i_{4p-1}$, gives condition (\ref{fourfour}) if the SSNB is zero, 
{\it q.e.d.}

\section{Linear generalized super-Poisson structures on the duals of simple 
Lie superalgebras}

Given a finite-dimensional Lie superalgebra $\cal G$, we know from Sec. III 
that there is a linear super-Poisson structure defined through the structure
constants. If $\cal G$ admits a non-degenerate Killing metric
$k_{ij}$, one may, on the other hand, construct the graded
skew-symmetric order three tensor 
\be
\omega(e_i,e_j,e_k)
:=k([e_i,e_j],e_k)=C_{ij}^{l}k_{lk}=C_{ijk}\,,\, 
e_i\in\g\quad (i,j,k=1,\ldots,r=\hbox{dim}\g)
\label{VIi}
\ee
where the $e_i$ are elements of a basis of $\cal G$.
This tensor is invariant, \ie
\be
\omega([e_l,e_i],e_j,e_k)+(-1)^{\alpha(l)\alpha(i)}\omega(e_i,[e_l,e_j],e_k)+
(-1)^{[\alpha(i)+\alpha(j)]\alpha(l)}\omega(e_i,e_j,[e_l,e_k])=0
\quad.
\label{VIii}
\ee
In general, the invariance (or $ad$-invariance) of a tensor of components 
$k_{i_1...i_m}$ may be expressed as
\be
    \sum_{s=1}^m (-1)^{\alpha(j)[\alpha(i_1)+...+\alpha(i_{s-1})]}C^\rho_{ji_s}
     k_{i_1...i_{s-1}\rho i_{s+1}...i_m}=0\quad ,        \label{VIiii}
\ee
which for the case of a graded skew-symmetric tensor $\omega$ can be written as
\be
    {\tilde \epsilon}^{j_1...j_m}_{i_1...i_m}C^\rho_{kj_1}
      \omega_{\rho j_2...j_m}=0\quad .                  \label{VIiv}
\ee
Since we are assuming that $k$ is non-degenerate, it can be used to raise indices as well,
so starting from $C_{ijk}$ as defined in (\ref{VIi}), one can recover
the structure constants and the corresponding super-Poisson structure.
This fact was used in \cite{APPa} to obtain linear generalized Poisson 
structures from simple Lie algebras by using certain invariant skew-symmetric 
forms of odd order (Lie algebra cohomology cocycles). 
These forms were obtained starting from $ad$-invariant
symmetric polynomials (Casimirs), which are completely classified for
simple Lie algebras. 
However, in contrast with this case, the ring of Casimir operators for simple 
superalgebras (the center $Z(U(\g))$ of the enveloping algebra) is not finitely 
generated in general
(among the classical superalgebras this is the case only for 
$osp(1,2n)$ for which $Z(U(\g))$ is generated by $n$ Casimir operators of 
order $2,4,\ldots,2n$). 
At the same time, the study of the invariant 
polynomials for superalgebras is much more involved than for the ordinary 
simple Lie algebras case (see in this respect 
\cite{JarGreen,ANS,Kacii,Berezin,Scheunert} 
and references therein). Also, there is the problem that for
simple superalgebras the Killing form may be zero since the invariance and 
simplicity entails that $k$ is either non-degenerate or identically zero ($k$ 
is non-degenerate for the following classical superalgebras: 
$A(m,n)\,,\,m>n\ge 0\ [sl(m+1,n+1)];\ B(m,n)\,,\,
m\ge 0\,,\,n\ge 1\ [osp(2m+1,2n)];\ C(n)\,,\,n\ge 2\ [osp(2,2n-2)];\
D(m,n)\,,\,m\ge 2\,,\,n\ge 1\,,\,m\ne n+1\ [osp(2m,2n)];\ F(4)$ and $G(3)$ 
\cite{Kac}; see also \cite{dW}). 
We shall assume here that the Killing form is non-degenerate and consider 
Casimir operators defined by $ad$-invariant supersymmetric polynomials.
We shall describe now how to obtain linear super-Poisson structures in this 
case.

\medskip
\noindent
{\bf Theorem V.1} ({\it Linear generalized SPS on a simple superalgebra})

\noindent Let $\g$ be a simple superalgebra, and let 
$k_{i_1\ldots i_m}$ be
a primitive non-trivial invariant graded-symmetric polynomial of order $m$.
Then, the tensor $\omega_{\rho l_2\ldots l_{2m-2} \sigma}$

\be
\omega_{\rho l_2\ldots l_{2m-2} \sigma}:=
{\tilde \epsilon}^{j_2\ldots j_{2m-2}}_{l_2\ldots l_{2m-2}}
\tilde\omega_{\rho j_2\ldots j_{2m-2} \sigma}
\;,\;
\tilde\omega_{\rho j_2\ldots j_{2m-2}\sigma}:=
k_{\rho i_1\ldots i_{m-1}}
C^{i_1}_{j_2 j_3}\ldots C^{i_{m-1}}_{j_{2m-2}\sigma}
\label{fiveone}
\ee
is completely graded skew-symmetric and 
\be
\Lambda^{(2m-2)}={(-1)^{m-1}\over (2m-2)!}x_\sigma 
\omega_{\cdot l_1\ldots l_{2m-2}}^\sigma
\partial^{l_{2m-2}}\wedge\ldots\wedge \partial^{l_{1}}
\label{fivetwo}
\ee
defines a linear generalized super-Poisson structure on $\g$ by
\be
\{x_{i_1},\dots,x_{i_{2m-2}}\}=
x_\sigma \omega_{\cdot i_1\ldots i_{2m-2}}^\sigma\quad.
\label{defines}
\ee
\noindent
{\it Proof:\quad} 
Let us first consider the complete graded skew-symmetry.
Since $\omega_{\rho l_2\ldots l_{2m-2} \sigma}$ is, by (\ref{fiveone}), graded 
skew-symmetric under the interchange of
$l_i$, $l_j$ $i,j=2,...,2m-2$ and under the interchange of $\sigma$ and $l_i$,
it suffices to prove the graded skew-symmetry relative to the indices
$\rho$ and $\sigma$. This can be done by using the $ad$-invariance of
$k$ (eq. (\ref{VIiii})) to rewrite (\ref{fiveone}) as
\be
\eqalign{
\omega_{\rho l_2\ldots l_{2m-2} \sigma}=
&-(-1)^{\alpha(j_{2m-2})[\alpha(\rho)+\alpha(i_1)+...+\alpha(i_{m-2})]}
{\tilde \epsilon}^{j_2\ldots j_{2m-2}}_{l_2\ldots l_{2m-2}}\cr
&
\hskip -30pt
\big[ \sum_{s=1}^{m-2}(-1)^{\alpha(j_{2m-2})[\alpha(\rho)+\alpha(i_1)+...
+\alpha(i_{s-1})]} k_{\rho i_1...i_{s-1}i_{m-1}i_{s+1}...i_{m-2}\sigma}
C^{i_1}_{j_2 j_3}\ldots C^{i_{m-1}}_{j_{2m-2}i_s}\cr
&
\hskip -30pt
+k_{i_{m-1} i_1...i_{m-2}\sigma}
C^{i_1}_{j_2 j_3}\ldots C^{i_{m-1}}_{j_{2m-2}\rho}\big]  \quad .\cr}
                                                        \label{fivetwoa} 
\ee 
By using that $\alpha(i)=\alpha(j)+\alpha(k)$ if $i,j,k$ are the indices of a 
$C_{jk}^i$ commutator and the graded
skew-symmetry and symmetry properties of $\tilde\epsilon$ and $k$ respectively,
it is easily seen that the first term is equal to
\be
    \eqalign{
    \sum_{s=1}^{m-2}
{\tilde \epsilon}^{j_2...j_{2s}j_{2s+1}j_{2m-2}j_{2s+2}...
j_{2m-3}}_{l_2...l_{2m-2}}
& k_{\rho i_1...i_{s-1}i_{m-1}i_{s+1}...i_{m-2}\sigma}\cr
&    \cdot 
  C^{i_1}_{j_2 j_3}...C^{i_s}_{j_{2s} j_{2s+1}}C^{i_{m-1}}_{i_s j_{2m-2}}
  C^{i_{s+1}}_{j_{2s+2} j_{2s+3}}...C^{i_{m-2}}_{j_{2m-4} j_{2m-3}}\quad ,\cr}
                                                         \label{fivetwob}
\ee
which is zero due to the ordinary super-Jacobi identity involving
$C^{i_s}_{j_{2s} j_{2s-1}}\allowbreak C^{i_{m-1}}_{i_s j_{2m-2}}$.
Thus, $\omega$ reduces to the second term in (\ref{fivetwoa}) and reads
\be
   \eqalign{
    \omega_{\rho l_2\ldots l_{2m-2}\sigma}=&
    -(-1)^{\alpha(\rho)\alpha(\sigma)+[\alpha(\rho)+\alpha(\sigma)]
    [\alpha(j_2)+...+\alpha(j_{2m-2})]}\cr
    &{\tilde \epsilon}^{j_2\ldots 
    j_{2m-2}}_{l_2\ldots l_{2m-2}}k_{\sigma i_1...i_{m-1}}
    C^{i_1}_{j_2 j_3}\ldots C^{i_{m-1}}_{j_{2m-2}\rho}\cr
    =&-(-1)^{\alpha(\rho)\alpha(\sigma)+[\alpha(\rho)+\alpha(\sigma)]
    [\alpha(l_2)+...+\alpha(l_{2m-2})]}\omega_{\sigma l_2\ldots l_{2m-2}\rho}
                \quad , \cr}                               \label{fivetwoc}
\ee
where the last equality is due to the fact that the presence of
$\tilde\epsilon$ means that $\alpha(j_2)+...+\alpha(j_{2m-2})=\alpha(l_2)+...
+\alpha(l_{2m-2})$.
Hence, $\omega$ is graded skew-symmetric.

Due to Lemma IV.1, the second part of the theorem requires checking the 
generalized super-Jacobi
identity for $\{ x_{i_1},...,x_{i_{2m-2}} \}=x_\sigma
{\omega_\cdot^\sigma}_{i_1...i_{2m-2}}$, which means computing
\be
\eqalign{
&    {\tilde\epsilon}^{i_1...i_{4m-5}}_{j_1...j_{4m-5}}x_\sigma
     {\omega_\cdot^\sigma}_{i_1...i_{2m-3}\rho}
     {\omega_\cdot^\rho}_{i_{2m-2}...i_{4m-5}} \cr
&    =(2m-3)! {\tilde\epsilon}^{i_1...i_{4m-5}}_{j_1...j_{4m-5}}x_\sigma
     {k^\sigma}_{l_1...l_{m-1}}
     C^{l_1}_{i_1 i_2}\ldots C^{l_{m-1}}_{i_{2m-3}\rho}{\omega^\rho}_{i_{2m-2}
     ...i_{4m-5}}\cr
&     
=(2m-3)! (-1)^{\alpha(l_{m-1})}{\tilde\epsilon}^{i_1,...,i_{4m-5}}_{j_1,...,
     j_{4m-5}}x_\sigma
     {{k^\sigma}_{l_1...l_{m-2}}}^{l_{m-1}}
     C^{l_1}_{i_1 i_2}\ldots C^{\rho}_{l_{m-1}i_{2m-3}}{\omega}_{\rho i_{2m-2}
     ...i_{4m-5}}\,,\cr}                        \label{fivetwod}
\ee
where we have used  (\ref{fiveone}) for one of the two $\omega$ factors, that 
$U^i V_i=(-1)^{\alpha(i)}U_i V^i$ (which follows from the graded symmetry of 
the Killing matrix) and eq. (\ref{VIi}). 
It is clear from
(\ref{fivetwod}) and (\ref{VIiv}) that the generalized super-Jacobi identity is 
satisfied if $\omega$ is $ad$-invariant. But this is indeed the case, due 
to the $ad$-invariance of $k$: substituting (\ref{fiveone}) in the left
hand side of (\ref{VIiv})
and then using again (\ref{VIiii}),  
\be
\eqalign{ 
{\tilde \epsilon}_{i_1...i_{2m-1}}^{j_1...j_{2m-1}}
    C^k_{\alpha j_1} \omega_{k j_2...j_{2m-1}}
      =&(2m-3)!{\tilde \epsilon}_{i_1...i_{2m-1}}^{j_1...j_{2m-1}}
      \big( \sum_{s=2}^{m} (-1)^{\alpha(j_1)[\alpha(l_2)+
      ...+\alpha(l_{s-1})]}\cr
      &-k_{k l_2...l_{s-1} l_1 l_{s+1}...l_m}
      C^{l_1}_{j_1l_2}C^{l_2}_{j_2j_3}...C^{l_m}_{j_{2m-2}j_{2m-3}}\big)
      =0 \cr}                                           \label{fivetwoe}
\ee
which is easily seen to vanish by bringing the index $j_1$ next to
$j_{2s-1}$, with the corresponding sign from $\tilde\epsilon$, and then using
the ordinary super-Jacobi identity, {\it q.e.d}

In fact, it may be shown that different 
$\Lambda^{(2m-2)}\,,\,\Lambda^{(2m'-2)}$ tensors also commute with 
respect to the SSNB and that they are functionally independent.

In practice, given a matrix representation $X_i$ of the superalgebra 
$\cal G$, the supertraces (see, \eg\ \cite{dW}) of the graded-symmetric 
product of several generators define invariant polynomials \ie,
\be
     k_{i_1...i_m}\propto \hbox{ssTr}(X_{i_1}...X_{i_m})\quad , \label{fivethree}
\ee
where ssTr means graded-symmetric supertrace and of which the Killing form 
$k_{ij}$=$\hbox{sTr}(ad\,X_i\,\allowbreak ad\,X_j)$ is the lowest order example.
The fact that these tensors are invariant is deduced easily from
the `cyclic' property of the supertrace, sTr$(AB)=(-1)^{\alpha(A)\alpha(B)}
\hbox{sTr}(BA)$. Indeed, using the definition (\ref{VIiii}),
\be
     \eqalign{\sum^m_{s=1} &(-1)^{\alpha(j)[\alpha(i_1)+...+\alpha(i_{s-1})]}
   C^\rho_{ji_s}k_{i_1...i_{s-1}\rho i_{s+1}...i_m}\cr
    &=\hbox{sTr}\big(\sum^m_{s=1} (-1)^{\alpha(j)[\alpha(i_1)+...+\alpha(i_{s-1})]}
    X_{i_1}...X_{i_{s-1}}[X_{j},X_{i_s}]X_{i_{s+1}}...X_{i_m}\big)\cr
    &=\hbox{sTr}\big([X_j,X_{i_1}]X_{i_2}...X_{i_m}+(-1)^{\alpha(j)\alpha(i_1)}
   X_{i_1}[X_j,X_{i_2}]X_{i_3}...X_{i_m}\cr
   &+...+(-1)^{\alpha(j)[\alpha(i_1)+...+\alpha(i_{m-1})]}
   X_{i_1}...X_{i_{m-1}}[X_j,X_{i_m}]\big)\cr
   &=\hbox{sTr}(X_jX_{i_1}...X_{i_m}-(-1)^{\alpha(j)[\alpha(i_1)+...
     +\alpha(i_m)]}X_{i_1}...X_{i_m}X_j)=0\quad .  \cr}\label{fivefour}
\ee

\section{An example of generalized super-Poisson structure}

As we have seen, the construction of a linear SPS following the procedure 
of Sec. V uses a graded-symmetric invariant polynomial on a simple superalgebra 
with non-degenerate Killing form.
This does not always exist: for certain simple Lie superalgebras
$k$ is identically zero \cite{Kac}. 
An example that does not present this problem is $su(3,1)$, the simplest 
superunitary algebra containing $su(3)$. 
A simpler example would be the unitary orthosymplectic superalgebra $uosp(2,1)$, 
which contains $su(2)$, but it does not have primitive graded-symmetric 
polynomials of order higher than two (much in the same way $su(2)$, being of 
rank one, has only one primitive Casimir operator) and for it eq. (\ref{fiveone})
reduces to the SPS given by the $\omega$ in (\ref{IVvii}).

Instead of making all the structure constants and commutators/anticommutators 
of the rank three, fifteen-generator $su(3,1)$-superalgebra explicit, we shall 
proceed in a more basic way which will allow us to
exhibit the essentials of our general procedure.
To this aim, consider first the identity
\be
\epsilon^{ijkl}_{1234}=
\epsilon^{ij}_{12}\epsilon^{kl}_{34}-\epsilon^{ij}_{13}\epsilon^{kl}_{24}+
\epsilon^{ij}_{14}\epsilon^{kl}_{23}+\epsilon^{ij}_{23}\epsilon^{kl}_{14}
-\epsilon^{ij}_{24}\epsilon^{kl}_{13}+\epsilon^{ij}_{34}\epsilon^{kl}_{12}
\quad,
\label{example1}
\ee
for the standard Levi-Civita tensor.
This means that the 24 terms in the four-commutator $[X_1,X_2,X_3,X_4]$, which 
is defined as the antisymmetrized sum of all products of the four generators, 
may be expressed as a sum of three terms
\be
[X_1,X_2,X_3,X_4]=\{[X_1,X_2],[X_3,X_4]\}+
\{[X_2,X_3],[X_1,X_4]\}
+\{[X_3,X_1],[X_2,X_4]\}
\quad,
\label{example2}
\ee
where $[\ ,\ ]$ ($\{\ ,\ \}$) means commutator (anticommutator). 
We may easily extend this to the $Z_2$-graded case.
If we now use the $Z_2$-graded commutator with
\be
[X,Y]:=XY-(-1)^{\alpha(X)\alpha(Y)}YX \quad (=-(-1)^{\alpha(X)\alpha(Y)}[Y,X])
\label{example3}
\ee
(\ie, the SSNB (\ref{ssnba}) for the elements of a superalgebra which reduces to 
a commutator or, in the odd/odd case, to the anticommutator), relation 
(\ref{example2}) above becomes in the graded case 
\be
\eqalign{
[X_i,X_j,X_k,X_l]=&\{[X_i,X_j],[X_k,X_l]\}+
(-1)^{\alpha(i)\alpha(k)+\alpha(i)\alpha(j)}\{[X_j,X_k],[X_i,X_l]\}\cr
&+(-1)^{\alpha(i)\alpha(k)+\alpha(j)\alpha(k)}\{[X_k,X_i],[X_j,X_l]\}
\quad,
\cr}
\label{example4}
\ee
where $\{\ ,\ \}$ is now the $Z_2$-graded {\it anticommutator} defined by
\be
\{X,Y\}:=XY+(-1)^{\alpha(X)\alpha(Y)}YX \quad 
(=(-1)^{\alpha(X)\alpha(Y)}\{Y,X\})
\quad.
\label{example5}
\ee
Expression (\ref{example4}) of course follows from the equivalent to 
(\ref{example1}) 
for the $Z_2$-graded case \ie,
\be
\eqalign{
\tilde\epsilon^{ijkl}_{1234}=&
\tilde\epsilon^{ij}_{12}\tilde\epsilon^{kl}_{34}
-(-1)^{\alpha(2)\alpha(3)}\tilde\epsilon^{ij}_{13}\tilde\epsilon^{kl}_{24}
\cr &
+(-1)^{\alpha(4)(\alpha(2)+\alpha(3))}
\tilde\epsilon^{ij}_{14}\tilde\epsilon^{kl}_{23}+
(-1)^{\alpha(1)(\alpha(2)+\alpha(3))}
\tilde\epsilon^{ij}_{23}\tilde\epsilon^{kl}_{14}
\cr &
-(-1)^{\alpha(1)\alpha(2)+\alpha(4)(\alpha(1)+\alpha(3))}
\tilde\epsilon^{ij}_{24}\tilde\epsilon^{kl}_{13}+
(-1)^{(\alpha(3)+\alpha(4))(\alpha(1)+\alpha(2))}
\tilde\epsilon^{ij}_{34}\tilde\epsilon^{kl}_{12}\quad.
\cr}
\label{example6}
\ee

Let us write the graded commutators among the elements of a 
basis of $su(3,1)$ as $[X_a,X_b]=C_{ab}^cX_c$.
For the graded anticommutators (\ref{example5}) we have, in {\it this} case, 
the generic form
\be
\{X_a,X_b\}=k_{ab}^c X_c + \delta_{ab} I\quad,
\label{example7}
\ee
where $I$ represents a central element.
This means that the {\it r.h.s} of (\ref{example4}) above may be written as
\be
\eqalign{
C_{ij}^mC_{kl}^n(k_{mn}^p X_p+\delta_{mn})
&+(-1)^{\alpha(i)\alpha(k)+\alpha(i)\alpha(j)}
C_{jk}^mC_{il}^n(k_{mn}^p X_p+\delta_{mn})\cr &
+(-1)^{\alpha(i)\alpha(k)+\alpha(j)\alpha(k)}
C_{ki}^mC_{jl}^n(k_{mn}^p X_p+\delta_{mn})
\quad.
\cr}
\label{example8}
\ee
It is easy to see that terms with $\delta$ cancel by using the graded Jacobi 
identity (\ref{VIa}) for the structure constants.
Grouping the terms in $k$, we may see that
\be
[X_i,X_j,X_k,X_l]={1\over 2}
\tilde\epsilon_{ijk}^{rst}C_{rs}^mC_{tl}^nk_{mn}^p X_p=
{1\over 2} \omega_{ijkl}^p X_p\quad,
\label{example9}
\ee
where the $\omega$ appearing here has, precisely, the structure of (\ref{fiveone}).
Using this result, it is clear that 
$\{x_{i_1},x_{i_2},x_{i_3},x_{i_4}\}=x_\sigma \omega_{i_1 i_2 i_3 i_4}^\sigma$ 
(see (\ref{defines})) defines a linear GSPS on $su(3,1)^*$.
Note that this depends on the existence of the graded symmetric polynomial 
$k_{ab}^c$ in (\ref{example7}), which is a specific property of $su(3,1)$.
However, the procedure may be extended to other algebras along similar lines. 


\section*{Acknowledgements}
This research has been partially supported by the CICYT and the DGICYT, Spain 
(AEN 96-1669, PR 95-439). 
J.M.I thanks the HCM programme of
the European Union for financial support;
A.P. thanks the Vice-rectorate of research of Valencia University for 
making his stay in Valencia possible. 
J.A. and J.C.P.B. wish to acknowledge the kind hospitality extended to them 
at DAMTP; they also wish to thank St. John's College for support (J.A.) and
the Spanish Ministry of Education and Science and the CSIC for an FPI grant 
(J.C.P.B.).



\begin{thebibliography}{99}

\bibitem{APPa}
{J.A. de Azc\'arraga, A.M. Perelomov and J.C. P\'erez Bueno, 
{\it New generalized Poisson structures}, J. Phys. {\bf A29}, L151-L157
(1996);
{\it The Schouten-Nijenhuis bracket, cohomology and generalized 
Poisson structures}, J. Phys. {\bf A29}, 7993-8009 (1996)}

\bibitem{Na}{
Y. Nambu, 
{\it Generalized Hamiltonian dynamics},
Phys. Rev. {\bf D7}, 2405--2412 (1973)}

\bibitem{BF}
{F. Bayen and M. Flato, 
{\it Remarks concerning Nambu's generalized mechanics}, 
Phys. Rev. {\bf D11}, 3049--3053 (1975)}

\bibitem{MS}
{N. Mukunda and E. Sudarshan,
{\it Relation between Nambu and Hamiltonian mechanics}, 
Phys. Rev. {\bf D13}, 2846--2850 (1976)}

\bibitem{Ta}{L. Takhtajan,
{\it On foundations of the generalized Nambu mechanics},
Commun. Math. Phys. {\bf 160}, 295--315 (1994)}

\bibitem{NOS}{J.A. de Azc\'arraga, J.M. Izquierdo and J.C. P\'erez Bueno,
{\it On the generalizations of Poisson structures}, DAMTP 97-12,
hep-th/9703019}

\bibitem{Sc}{J.A. Schouten,
{\it \"Uber Differentialkonkomitanten zweier kontravarianter Gr\"ossen}, 
Proc. Kon. Ned. Akad. Wet. Amsterdam {\bf 43}, 449-452 (1940)}

\bibitem{Ni}{A. Nijenhuis,
{\it Jacobi-type identities for bilinear differential concomitants 
of certain tensor fields}, 
Indag. Math. {\bf 17}, 390-403 (1955)}

\bibitem{Lich}{A. Lichnerowicz,
{\it Les vari\'et\'es de Poisson et leurs alg\`ebres de Lie associ\'ees},
J. Diff. Geom. {\bf 12}, 253-300 (1977)} 

\bibitem{Tu}{W.M. Tulczyjew,
{\it Poisson brackets and canonical manifolds}, 
Bull. Acad. Pol. Sci. (Math. and Astronomy) {\bf 22}, 931--934 (1974)}

\bibitem{Leites}
{D.A. Leites, {\it New superalgebras and mechanics},
Sov. Math. Dokl. {\bf 18}, 1277-1280 (1977)}

\bibitem{Kuper}
{B. Kupershmidt, {\it Odd and even Poisson brackets in dynamical systems}, 
Lett. Math. Phys. {\bf 9}, 323-330 (1985)}

\bibitem{JM}
{J. Monterde, {\it A characterization of graded symplectic structures},
Diff. J. Geom. and Appl. {\bf 2}, 81-97 (1992)}

\bibitem{JG}
{J. Grabowski, {\it Graded extensions of Poisson brackets}, talk at the Goslar 
{\it Int. Coll. on Group Theor. Methods in Phys.}, July 1996}

\bibitem{Kras}
{I.S. Krasil'shchik,
{\it Supercanonical algebras and Schouten bracket},
Mat. Zametki {\bf 49}, 70-76 (1991) (Math. Notes {\bf 49}, 50-54 (1991))}

\bibitem{CI}{F. Cantrijn, and A. Ibort, 
{\it Introduction to Poisson supermanifolds}, Diff. Geom. and
Appl. {\bf 1}, 133-152 (1991)}

\bibitem{Rogers}{A. Rogers, {\it A global theory of supermanifolds}, 
J. Math. Phys. {\bf 21}, 1352-1365 (1980)}

\bibitem{dW}{B. De Witt, 
{\it Supermanifolds}, 2nd ed. Camb. Univ. Press (1992)}

\bibitem{Berezin}{F.A. Berezin, 
{\it Introduction to superanalysis}, D. Reidel, Dordrecht (1987)}

\bibitem{Kostant}{B. Kostant, 
{\it Graded manifolds, graded Lie theory and prequantization}, 
in Lect. Notes in Math. {\bf 570}, Springer-Verlag (1977), p. 177-306}

\bibitem{Batchelor}{M. Batchelor, {\it Two approaches to supermanifolds},
Trans. Am. Math. Soc. {\bf 258}, 257-270 (1980)}

\bibitem{KS}{B. Kostant and S. Sternberg, {\it Symplectic reduction, BRST
cohomology and infinite dimensional Clifford algebras}, Ann. Phys. {\bf 176},
49-113 (1987)}

\bibitem{FN}
{A. Fr\"olicher and A. Nijenhuis, {\it Theory of vector-valued differential 
forms}, Indag. Math. {\bf 18}, 338-350; 351-359 (1956)}

\bibitem{YK}
{Y. Kosmann-Schwarzbach, 
{\it From Poisson algebras to Gerstenhaber algebras},
Ann. Inst. Fourier {\bf 46}, 1243-1274 (1996)}

\bibitem{LMS}
{P.A.B. Lecomte, P.W. Michor and H. Schicketanz, 
{\it The multigraded Nijenhuis-Richardson algebra, its universal property and 
its applications}, J. Pure and Applied Alg. {\bf 77}, 87-102 (1992)}

\bibitem{CE}{C. Chevalley and S. Eilenberg, {\it Cohomology theory of
Lie groups and Lie algebras}, Trans. Am. Math. Soc. {\bf 63}, 85--124 
(1948)} 

\bibitem{AI}{J.A. de Azc\'arraga and J.M. Izquierdo, {\it Lie groups, Lie 
algebras, cohomology and some applications in physics}, Camb. Univ. Press 
(1995)}

\bibitem{Kac}{V.G. Kac, {\it Lie superalgebras},
Adv. in Math. {\bf 26}, 8-96 (1977); {\it A Sketch of Lie Superalgebra Theory},
Commun. Math. Phys. {\bf 53}, 31-64 (1977)}

\bibitem{SCHEUNERT}
{M. Scheunert, {\it The theory of Lie superalgebras}, Springer Verlag (1979)}

\bibitem{Buttin}
{C. Buttin, 
{\it Les d\'erivations des champs de tenseurs et l'invariant diff\'erentiel de 
Schouten}, 
C.R. Acad. Sci. Paris {\bf 269A}, 87-89 (1969)}

\bibitem{BV}{I.A. Batalin and G.A. Vilkovisky,
{\it Quantization of gauge theories with linearly dependent generators}, 
Phys. Rev. {\bf D28}, 2567 (1984) (Errata: {\bf D30}, 508 (1984)).
For a review see J. Gomis, J. Paris and S. Samuel, 
{\it Antibrackets, antifields and field quantization},
Phys. Rep. {\bf 259}, 1-145 (1995)}

\bibitem{Witten}
{E. Witten, 
{\it A note on the antibracket formalism}, Mod. Phys. Lett. {\bf A5}, 487-494 
(1990)}

\bibitem{LZ}
{B.H. Lian and G.J. Zuckerman, 
{\it New perspectives on the BRST-algebraic structure of string theory},
Commun. Math. Phys. {\bf 154}, 613-646 (1993)}

\bibitem{ZWI}
{B. Zwiebach,
{\it Closed string theory: quantum action and the Batalin-Vilkovisky master
equation},
Nucl. Phys. {\bf B390}, 33-152 (1993)}

\bibitem{AS}
{A. Schwarz,
{\it Geometry of Batalin-Vilkovisky quantization},
Commun. Math. Phys. {\bf 155}, 249-260 (1993);
M. Alexandrov, M. Kontsevich, A. Schwarz and O. Zaboronsky,
{\it The geometry of the master equation and topological quantum field 
theory}, to appear in Int. J. Mod. Phys.}

\bibitem{HIGHER}
{J.A. de Azc\'arraga and J.C. P\'erez Bueno,
{\it Higher-order simple Lie algebras}
(hep-th/9605213), to appear in Commun. Math. Phys.}

\bibitem{JarGreen}
{P.D. Jarvis and H.S. Green, 
{\it Casimir invariants and characteristic identities for generators of the 
general linear, special linear and orthosymplectic graded Lie algebras},
J. Math. Phys. {\bf 20}, 2115-2122 (1979)}

\bibitem{ANS}
{A. N. Sergeev, C. R. Acad. Bulgare Sci. {\bf 35}, 573 (1982);
{\it Analogue of the classical invariant theory for Lie superalgebras}, Func. 
Anal. and Appl. {\bf 26}, 223-225 (1992)}

\bibitem{Kacii}
{V. G. Kac, {\it Laplace operators
of infinite-dimensional Lie algebras and theta functions},
Proc. Nat. Acad. Sci. USA {\bf 81}, 645-647 (1984)}

\bibitem{Scheunert}{M. Scheunert,
{\it Casimir elements of Lie superalgebras} in S. Sternberg (ed.), 
in {\it Differential geometric methods in Math. Physics}, p. 115-124, 
D. Reidel (1984);
{\it Invariant supersymmetric multilinear 
forms and the Casimir elements of $P$-type Lie superalgebras}, J. Math.
Phys. {\bf 28}, 1180-1191 (1987) }

\end{thebibliography}
\end{document}